\documentclass[12pt]{RR-article}
\usepackage{amsmath,amsthm,statmath,amssymb,bm,graphicx}
\usepackage{epsfig}
\usepackage{indentfirst}
\usepackage{caption}
\usepackage{subcaption}
\usepackage[backend=biber, style=apa, citestyle=apa]{biblatex}
\usepackage{setspace}
\usepackage{float}

\begin{document}
\newcommand{\comment}[1]{}
\newcommand{\bs}{\boldsymbol}
\newcommand{\N}{\mathcal{N}}
\newcommand{\Cov}{\mathop{\mathrm{Cov}}\nolimits}
\newcommand{\Var}{\mathop{\mathrm{Var}}\nolimits}
\newcommand{\bas}{\renewcommand{\baselinestretch}}
\newtheorem{theorem}{Theorem}
\renewcommand{\baselinestretch}{1.5}
\title{\large{\bf Corrected Version of ``Simplicity Above Elegance: Another Look at the Asymptotically Correct Standardization of Snijders''}}
\author{Sandip Sinharay, ETS}
\date{\today}
\vspace{40cm} \maketitle \pagestyle{plain} \thispagestyle{empty}

NOTE: A previous version of this article has been published as\\
Sinharay, S. (2026). Simplicity above elegance: Another look at the asymptotically correct standardization of Snijders. {\it Journal of Educational and Behavioral Statistics.} https://doi.org/10.3102/10769986251415200.\\
After publication of the above, an error was found in it, which was published in a corrigendum in the same journal. This version is the corrected version of the article, that is, this version implements the correction made in the corrigendum. Due to the correction, the content  in pages 7-12 of this document differs substantially from that in the journal version.
\newpage
\setlength{\topmargin}{0in} \pagenumbering{roman}
\begin{center}
{\bf Simplicity Above Elegance: Another Look at the Asymptotically Correct Standardization of Snijders}
\end{center}

\centerline{\bf Abstract} \noindent
Person-fit statistics are widely used to detect aberrant response patterns
in educational and psychological measurement. \textcite{s01} suggested an 
 asymptotically correct standardization for a broad class of such statistics. 
This paper presents an alternative derivation of this standardization. The derivation yields several advantages including a simpler formula and simpler description  of several person-fit statistics including the $l^*_z$ statistic~\parencite{vm99} and a theoretical explanation of simulation findings reported by \textcite{s01} and \textcite{vm99}, among others.

\vspace{1cm} \noindent{Key words}: Ability estimate; item response theory; $l_z$; $l^*_z$; person fit; Taylor series expansion; test fraud.
\newpage
\centerline {\bf Acknowledgements} \noindent
The author wishes to express sincere appreciation and gratitude to the editor Gongjun Xu and the two anonymous reviewers for their helpful comments. The author would also like to thank Sébastien Béland,  Ying ``Alison'' Cheng, Kylie Gorney, Tao Jiang, Matthew Johnson,  Zhongtian Lin, Frank Rijmen, Klaas Sijtsma, Jorge Tendeiro,  and Jiyun Zu for their helpful comments on earlier versions of this article. The author would also like to thank Prof. Tom Snijders for publishing his ``seminal'' paper, coincidentally, about 25 years ago. This paper would not exist without his paper.

\newpage
\pagenumbering{arabic} \setcounter{page}{1}
%
%
 Person-fit assessment concerns
 the detection of examinees with aberrant response patterns~\parencite[e.g.,][]{s01}. 
 \textcite{ws18} identified person-fit assessment as one of
  ﬁve statistical/psychometric approaches  used to detect fraud in educational tests.
  Person-fit assessment has also been applied for robust calibration~\parencite{hc19}, measuring person unreliability~\parencite{f14}, examining practical validity~\parencite{fvl16},
   identification of careless responses on a survey~\parencite[e.g.,][]{ms01},
   and screening of subtypes of suicide for clinicians by
identifying individuals who do not demonstrate typical response patterns \parencite[e.g.][]{hlc21}.

 Person-fit statistics constitute the primary statistical
 tool in person-fit assessment.
 The  $l_z$  statistic~\parencite{dlw85} is one of the most popular 
 person-fit statistics based on item response theory or IRT
 \parencite[e.g.,][]{t16}.  The statistic is a function of the examinee ability---so the computation of the statistic
 for a real data set requires an estimate of the examinee ability. 
 \textcite{s01} showed that the asymptotic null distribution (where ``null distribution'' refers to the distribution under no person misfit) of a broad class of  person-fit statistics including $l_z$, when computed with 
 an estimated examinee ability, is not standard normal as previously assumed by several users of the statistic.  \textcite{s01} also
 suggested an asymptotically correct standardization of these statistics  for dichotomous items. 
The purpose of this paper is to suggest an alternative  derivation of that  standardization. 
 
 The next section includes  a review of the standardization of  \textcite{s01}. 
 The Methods section includes an alternative derivation of the standardization. The connection between the standardized version of the person-fit statistics derived in this paper and the version of \textcite{s01} and the advantages of the derivation of this paper are also described in the section. The last section provides some conclusions and recommendations.
 
 All derivations involve the assumption that the IRT model is correct, that is, there 
 is no person misfit.\footnote{Consequently, for brevity, phrases like ``under no person misfit'' and ``when the IRT model fits'' will not appear in too many places from now on.} Because  \textcite{s01} considered only tests with dichotomous items, only those types of tests will be considered.
\section*{Review of the Asymptotically Correct Standardization of \textcite{s01}}
Consider an examinee with true ability  $\theta$  who was administered a 
  test that comprises $n$ dichotomous items. Let 
  $X_i$ denote the examinee's score (0 or 1)  on item $i$  and 
 $P_i(\theta)=P(X_i=1|\theta)$. In person-fit  assessment, the item parameters are assumed known and equal to their estimates---the same assumption will be made in this paper and hence the dependence of $P_i(\theta)$, $X_i$, and other mathematical expressions on item parameters will be suppressed in the notations. Also, as is typical in person-fit  assessment, $\theta$ will be assumed to be a fixed constant and the $X_i$'s for an examinee will be assumed independent of each other given $\theta$ due to the local independence assumption of IRT.

\textcite{s01} considered a  broad class of  person-fit statistics of the form
   \begin{eqnarray}
  Z(\theta)= \frac{W(\theta)-E(W(\theta))}{\sqrt{\Var(W(\theta))}},
 \label{Z} 
 \end{eqnarray}
where
  \begin{eqnarray}
 W(\theta) = \sum_{i=1}^{n} ( X_i - P_i(\theta) ) w_i(\theta)  
 \label{Wn} 
 \end{eqnarray}
 for suitable weight functions  $w_i(\theta)$ that do not depend on the $X_i$'s.
 From  Equation~\ref{Wn}, one obtains
 \begin{eqnarray}
E(W(\theta))=0 \mbox{ and } \Var(W(\theta))=n \sigma^2(\theta), \mbox{ where }\sigma^2(\theta)=\frac{1}{n}\sum_i  w^2_i(\theta) P_i(\theta)\{1-P_i(\theta)\}
 \label{ssq}
 \cdot
 \end{eqnarray}
 Consequently, $ Z(\theta)$  can be expressed as
   \begin{eqnarray}
  Z(\theta)=\frac{W(\theta)}{\sqrt{n} \sigma(\theta)} = \frac{W(\theta)}{\sqrt{\sum_i  w^2_i(\theta) P_i(\theta)\{1-P_i(\theta)\}}}
  \cdot
 \label{LzasWn} 
 \end{eqnarray}

  The statistic $ l(\theta) $, which is the log-likelihood of $\theta$ for the examinee, is defined as  \begin{eqnarray}
 l(\theta) = \sum_{i=1}^{n} \left \{ X_i \log P_i(\theta) + (1-X_i) \log \left( 1-P_i(\theta) \right) \right \} = \sum_{i=1}^{n} \left \{X_i \log \frac{P_i(\theta)}{1-P_i(\theta)} + \log( 1-P_i(\theta)) \right \}\cdot
  \label{l} 
 \end{eqnarray}
  The $l_z$ statistic for dichotomous items, which is defined as 
  \begin{eqnarray}
  l_z(\theta) = \frac{l(\theta)-E(l(\theta))}{\sqrt{\Var(l(\theta))}} = \frac{\sum_{i=1}^{n}  \left(X_i-P_i(\theta) \right)  \log \frac{P_i(\theta)}{1-P_i(\theta)} }{\sqrt{\Var(l(\theta))}}
 \label{lz} 
 \end{eqnarray}
 \parencite{dlw85}, is a special case of $Z(\theta)$ for  
 \begin{eqnarray}
 w_i(\theta)=\log\frac{P_i(\theta)}{1-P_i(\theta)},
 \label{wt}
 \end{eqnarray}
 as is clear from Equations~\ref{Wn}, \ref{LzasWn}, and \ref{lz}.
The infit and outfit mean squares \parencite{ws79,wm82} and two caution indices \parencite{t84}
 are also special cases  of the statistic~\parencite[e.g.][]{mbr14,s16APM}.

  The computation of a statistic of the form $Z(\theta)$ from a real data set requires
 the substitution of the unknown $\theta$  by an estimate.
 Let $\hat{\theta}$ denote such an estimate. 
  Researchers such as \textcite{mbr14}, \textcite{mh90}, \textcite{s01},  and \textcite{vm99}  found for simulated data that (a) the variance of the  null distribution of several special cases of $Z(\theta)$, when computed after substituting $\theta$ by $\hat{\theta}$, is smaller compared to 
  the standard normal distribution even for long tests,
 and, consequently, (b) the use of these statistics
 along with a standard normal null 
 distributional assumption leads to conservative person-fit assessment.

  \textcite{s01} proposed an asymptotically correct standardization of $Z(\theta)$. 
  The standardization is based on the first-order Taylor series approximation of $W(\hat{\theta})$ around $\theta$, which is given by
 \begin{eqnarray}
 \frac{1}{\sqrt{n}} W(\hat{\theta})  \approx  \frac{1}{\sqrt{n}}W(\theta) + \sqrt{n} (\hat{\theta}-\theta)
 \left \{   \frac{1}{n}\sum_i (X_i-P_i(\theta))w'_i(\theta) - \frac{1}{n} \sum_i P'_i(\theta)w_i(\theta)  \right \}, 
 \label{Taylor} 
 \end{eqnarray}
 where   $P'_i(\theta)$ and $w'_i(\theta)$  are the first derivatives
 of $P_i(\theta)$ and $w_i(\theta)$, respectively,  and 
 $\sqrt{n} (\hat{\theta}-\theta)$ is assumed to have a non-degenerate distribution as $n \rightarrow \infty$ so that it is  bounded. 
  Among the terms on the right-hand side  of Equation~\ref{Taylor},
 \begin{eqnarray}
 \frac{1}{\sqrt{n}}W(\theta) \indist \N(0,\sigma^2(\theta)) \mbox{ as } n \rightarrow \infty 
 \label{wtknown} 
 \end{eqnarray}
 \parencite{s01}, where the symbol $\indist$ means ``converges in distribution to''.
  The first term between the braces in Equation~\ref{Taylor}, $\frac{1}{n}\sum_i (X_i-P_i(\theta))w'_i(\theta)$, 
is an average of random variables 
 whose means are zero and hence will tend to 0 as $n$ increases.\footnote{The result holds because of the Chebyshev's theorem~\parencite[p. 112]{r73}, which is essentially the weak law of large numbers for non-iid random variables,  given the regularity condition assumed by \textcite{s01}   that $w'_i(\theta)$ is uniformly bounded.}
 Thus, for long tests, because $\sqrt{n} (\hat{\theta}-\theta)$ is bounded, 
 $\sqrt{n} (\hat{\theta}-\theta) \frac{1}{n}\sum_i (X_i-P_i(\theta))w'_i(\theta) \approx 0$, and hence 
 \begin{eqnarray}
 \frac{1}{\sqrt{n}} W(\hat{\theta})  \approx  \frac{1}{\sqrt{n}}W(\theta) - \sqrt{n} (\hat{\theta}-\theta)
 \left \{ \frac{1}{n} \sum_i P'_i(\theta)w_i(\theta)  \right \} \cdot
 \label{T21} 
 \end{eqnarray}
 The  term $\frac{1}{n}\sum_i P'_i(\theta)w_i(\theta)$, however, is not approximated well by 0 as $n$ increases. So, the asymptotic  distribution of $\frac{1}{\sqrt{n}}W(\hat{\theta})$ is not the same as that of $\frac{1}{\sqrt{n}}W({\theta})$, which means, given Equation~\ref{wtknown}, that
 the  asymptotic distribution of   $\frac{1}{\sqrt{n}}W(\hat{\theta})$ is not approximated well by a
 $ \N(0,\sigma^2(\theta))$  distribution. Thus,  for long tests, $Z(\hat{\theta})$  does not follow a standard normal distribution.
 
 \textcite{s01}  proved that if $\hat{\theta}$ satisfies the condition
   \begin{eqnarray}
  r_0(\hat{\theta})+\sum_{i=1}^{n} \left [ X_i - P_i(\hat{\theta})\right ] r_i(\hat{\theta})=0
 \label{Cond} 
 \end{eqnarray}
  for some $ r_0(\hat{\theta})$ and $r_i(\hat{\theta})$, and one defines
   \begin{eqnarray}
 \tilde{w}_i(\theta)= w_i(\theta) - c(\theta)r_i(\theta)
 \label{wtilde}
 \end{eqnarray}
   for
 \begin{eqnarray}
   c(\theta) = \frac{\sum_i P'_i(\theta) w_i(\theta)}
 {\sum_i P'_i(\theta) r_i(\theta)},
 \label{cn} 
 \end{eqnarray}
  then an asymptotically correct standardized version of ${Z}(\hat{\theta})$ is given by
 \begin{eqnarray}
  \tilde{Z}(\hat{\theta})=\frac{W(\hat{\theta})+c(\hat{\theta})r_0(\hat{\theta})}{\sqrt{n}\tau(\hat{\theta})},
  \label{Wdis}
  \end{eqnarray}
  for
   \begin{eqnarray}
   \tau^2(\theta) = \frac{1}{n} \sum_i \tilde{w}^2_i(\theta) P_i(\theta)\{1- P_i(\theta)\},
 \label{tau} 
 \end{eqnarray}
  and that $ \tilde{Z}(\hat{\theta})$ asymptotically follows the standard normal distribution
  under three mild regularity conditions.

   \textcite{s01} showed that his standardization result holds when $\hat{\theta}$
 is the  maximum likelihood estimate (MLE),
 weighted likelihood estimate \parencite[WLE,][]{w89},
  or modal a posteriori (MAP) estimate,  $r_i(\theta)$ given by
  \begin{eqnarray}
 r_i(\theta)=\frac{P'_i(\theta)}{P_i(\theta)\{1-P_i(\theta)\}},
 \label{ri}
 \end{eqnarray}
 $ r_0(\hat{\theta})$ is defined as 
 \[ r_0(\hat{\theta}) =
 \left\{ \begin{array}{l} 0  \mbox{ if } \hat{\theta}=\mbox{MLE},
   \\ \frac{d
     \log f(\hat{\theta})}{d\hat{\theta}} \mbox{ if }
   \hat{\theta}=\mbox{MAP}
  \\ 
\frac{J(\hat{\theta})}{2I(\hat{\theta})} \mbox{ if }
 \hat{\theta}=\mbox{WLE},
      \end{array}  
\right .\] where
 $f(\theta)$ is the prior distribution on $\theta$, 
 $J(\theta)=\sum_i  \frac{P'_{i}(\theta) P''_{i}(\theta)}{P_{i}(\theta)(1-P_{i}(\theta))}$,
  $P''_{i}(\theta)$  is the second derivative of $P_{i}({\theta})$,
 and the regularity conditions are satisfied for the common dichotomous IRT models.

 \textcite{s16BJ} proved that the results of \textcite{s01} hold when  
$\hat{\theta}$ is the posterior mean or expected a posteriori~(EAP) estimate~\parencite{bm82} for $ r_i(\theta)$ given by Equation~\ref{ri}
and also when $\hat{\theta}$ is 
a robust ability estimate like the biweight~\parencite{mb82} or Huber estimate~\parencite{sy11}.\footnote{The expression of $ r_i(\theta)$ for robust estimates is somewhat cumbersome and is given in Appendix~B.} The asymptotically correct standardized version of the $l_z$ statistic is popularly known as the $l^*_z$ statistic~\parencite[e.g.,][]{vm99} and has a standard normal asymptotic distribution.

Because of Equation~\ref{ri},  $c(\theta)$ provided in Equation~\ref{cn}
can be expressed as 
 \begin{eqnarray}
c(\theta)=\frac{\sum_i P'_i(\theta) w_i(\theta)}
 {\sum_i P'_i(\theta)r_i(\theta)}=  \frac{\sum_i P'_i(\theta) w_i(\theta)}
 {\sum_i \{P'_i(\theta)\}^2 /[ P_i(\theta)\{1- P_i(\theta)\}]} = \frac{1}{I(\theta)} \sum_i P'_i(\theta) w_i(\theta),  
  \label{cexpr}
  \end{eqnarray}
 because, for dichotomous IRT models,
 \begin{eqnarray}
 I(\theta)=\sum_i \frac{ \{P'_i(\theta)\}^2 }{P_i(\theta)\{1- P_i(\theta)\} }
 \label{It}
 \end{eqnarray}
\parencite[e.g.][p. 89]{hs85}.
\section*{Methods: An Alternative Derivation}
This section provides an alternative and arguably more direct derivation of the  standardization of \textcite{s01}. First,  an expression of the asymptotic variance of $W(\hat{\theta})$ is derived, and
  then, the asymptotic distribution of  $W(\hat{\theta})$ divided by the estimated square root of its approximate variance is derived.  
 \subsection*{The Assumptions}
 The derivations are predicated on the assumptions
 that $\hat{\theta}$ is 
 the MLE of ability or another ability 
 estimator that is asymptotically equivalent to (or can be approximated well, for long tests, by) the MLE\footnote{Note that this assumption implies that $\hat{\theta} \rightarrow \theta$ as $n \rightarrow \infty$.} and  that 
 \begin{eqnarray}
  \sqrt{n} (\hat{\theta}-\theta) \indist \N(0,\bar{I}^{-1}_0(\theta)),
   \label{thetaD}
 \end{eqnarray}
 where  $\bar{I}_0(\theta)$ is the (finite) limiting value, for $n\rightarrow \infty$, of
 $I(\theta)/n$, and 
 $I(\theta)$ is the Fisher information  for the examinee ability. 
 The assumption represented by Equation~\ref{thetaD} is similar to  the assumption of \textcite{s01} that $\sqrt{n} (\hat{\theta}-\theta)$ has a non-degenerate distribution as $n \rightarrow \infty$. 
It is also assumed, as in \textcite{s01}, that $P_i(\theta)$ and $w_i(\theta)$ are twice differentiable and these functions and their second derivatives are uniformly bounded in a neighborhood of $\theta$; these assumptions are satisfied for common dichotomous IRT models and several popular person-fit statistics.
\subsection*{Approximating $W(\hat{\theta})$ by a Function of $\theta$}
 Application of the first order Taylor-series expansion
 to $l'(\hat{\theta})$, which is the first derivative of $l(\theta)$ defined in Equation~\ref{l}, at $\theta=\hat{\theta}$, leads to the result that for long tests,
 \begin{eqnarray}
  l'(\hat{\theta}) \approx l'(\theta) + (\hat{\theta}-\theta)l''(\theta),
  \label{Tl}
 \end{eqnarray}
 where $l''(\theta)$ denotes the second derivative of $l(\theta)$, and hence 
 \begin{eqnarray}
 \hat{\theta} \approx \theta + \frac{1}{l''(\theta)}\{l'(\hat{\theta}) - l'(\theta) \} \cdot
 \label{l1} 
 \end{eqnarray}
  The derivative  $l'(\theta)$ is given by 
   \begin{eqnarray}
  l'(\theta) =  \sum_{i=1}^{n} \frac{X_i - P_i(\theta)}{P_i(\theta)\{1-P_i(\theta)\} } P'_i(\theta)
  \label{lprime}
  \end{eqnarray}
  \parencite[e.g.,][p. 82]{hs85}.
 Note that the MLE of ability is computed by solving the equation 
 $l'({\theta})=0$~\parencite[e.g.][p. 83]{hs85}---so $l'(\hat{\theta})$ is equal to 0 when $\hat{\theta}$ is the MLE. Because $\hat{\theta}$, when not equal to the MLE, is assumed to be
 asymptotically equivalent to the MLE,
  $l'(\hat{\theta}) \rightarrow 0$ as $n \rightarrow \infty $ for these estimates.
 So,  Equation~\ref{l1} leads to the result that for long tests, 
 \begin{eqnarray}
 \hat{\theta} \approx \theta + \frac{l'(\theta)}{-l''(\theta)} \cdot
 \label{thd} 
 \end{eqnarray} 
 This result is very similar to one provided in Equation~42 of \textcite{m16}.
 The quantity $-l''(\theta)$ in Equation~\ref{thd} is the observed information function for $\theta$ 
 and converges to its expected value, which is $I(\theta)$, the Fisher information  for the examinee ability, by the Chebyshev's theorem~\parencite[p. 112]{r73} given the aforementioned regularity conditions.\footnote{A detailed proof can be found in Lemma 1 of \textcite{m16} as applied to the MLE.} Therefore, Equation~\ref{thd} leads to the result that for long tests,
 \begin{eqnarray}
 \hat{\theta} - \theta \approx \frac{l'(\theta)}{I(\theta)} \cdot
 \label{thd2} 
 \end{eqnarray}
 Equations~\ref{Wn}, \ref{T21},   \ref{lprime}, and \ref{thd2} imply that 
\begin{eqnarray}
  W(\hat{\theta})  &\approx&  W(\theta) -  \frac{l'(\theta)}{I(\theta)}
 \left \{ \sum_i P'_i(\theta)w_i(\theta)  \right \} \nonumber \\
   &=&  \sum_i (X_i-P_i(\theta))w_i(\theta)   -  \frac{1}{I(\theta)}
 \left \{ \sum_i P'_i(\theta)w_i(\theta)  \right \} \sum_{i} \frac{X_i - P_i(\theta)}{P_i(\theta)\{1-P_i(\theta)\} } P'_i(\theta) \nonumber \\
 &=&  \sum_i (X_i-P_i(\theta))w_i(\theta)   -  c(\theta)  \sum_{i} \frac{X_i - P_i(\theta)}{P_i(\theta)\{1-P_i(\theta)\} } P'_i(\theta) \nonumber \\
  &=&  \sum_i (X_i-P_i(\theta))\left (w_i(\theta)   -  c(\theta)   \frac{P'_i(\theta)}{P_i(\theta)\{1-P_i(\theta)\} }  \right) \nonumber \\
  &=&  \sum_i (X_i-P_i(\theta)) (w_i(\theta)   -  c(\theta)   r_i(\theta)), 
 \label{T22} 
 \end{eqnarray}
 for  $r_i(\theta)$ and $c(\theta)$   provided in Equations~\ref{ri} and \ref{cexpr}, respectively. 
 Because of the local independence property of IRT models,  the terms $(X_i-P_i(\theta)) (w_i(\theta)   -  c(\theta)   r_i(\theta)), i=1,2,\ldots, n$, are independent of each other given $\theta$, and consequently, the variance of $W(\hat{\theta})$ can be approximated by the variance of the right-hand side of Equation~\ref{T22}  as
 \begin{eqnarray}
\Var(W(\hat{\theta})) &\approx&  \sum_i P_i(\theta)\{1-P_i(\theta)\}  (w_i(\theta)   -  c(\theta) r_i(\theta))^2 \nonumber\\
&=&\sum_i \tilde{w}^2_i(\theta) P_i(\theta)\{1-P_i(\theta)\} = n \tau^2(\theta) 
   \label{vw}
    \end{eqnarray}
 for expressions of $\tilde{w}_i(\theta) $ and $\tau^2(\theta)$ provided in  Equations~\ref{wtilde} and \ref{tau}.

  \subsection*{The Standardized Statistic and its Asymptotic Distribution}
 Define the standardized version of $W(\hat{\theta})$ as
 \begin{eqnarray}
 \tilde{\tilde{Z}}(\hat{\theta})=\frac{W(\hat{\theta})}{\sqrt{n}\tau(\hat{\theta})} \cdot
 \end{eqnarray}

  Let us denote the expression in the right-hand side of Equation~\ref{T22}, which is the sum of $n$ independent and non-identically distributed random variables, as $G(\theta)$.
   Because the  individual terms of $G(\theta)$ have mean of 0 and are uniformly bounded by the aforementioned regularity conditions, 
    \begin{eqnarray}
  \frac{G(\theta)}{  \sqrt{\Var( G(\theta) )} } = \frac{G(\theta)}{  \sqrt{n} \tau({\theta})  } \rightarrow  \N(0,1) 
  \label{gn}
    \end{eqnarray}
    by  Equation~\ref{vw} and the  Lindeberg-Feller central limit theorem for independent but   non-identical random variables~\parencite[e.g.,][p. 128]{r73}.
    Equations~\ref{T22}, \ref{vw}, and \ref{gn} imply that 
     \begin{eqnarray}
    \frac{W(\hat{\theta})}{ \sqrt{n} \tau({{\theta}})} \rightarrow \N(0,1) \cdot 
    \label{wn}
    \end{eqnarray} 
The standardized statistic $\tilde{\tilde{Z}}(\hat{\theta})$ can also be expressed as 
 \[ \tilde{\tilde{Z}}(\hat{\theta})=\frac{W(\hat{\theta})}{ \sqrt{n} \tau({{\theta}})}
 \frac{ \tau({{\theta}})}{  \tau({\hat{\theta}})}.
 \]
Because $\hat{\theta}$ converges to ${\theta}$, $ \frac{ \tau({{\theta}})}{  \tau({\hat{\theta}})} \rightarrow 1$
because of the result that $X_n \rightarrow c \Rightarrow g(X_n) \rightarrow g(c)$ for  any continuous function $g$ where $X_n$ is a sequence of random variables~\parencite[e.g.][p. 124]{r73}. 
Consequently,  because $\tilde{\tilde{Z}}(\hat{\theta})$ is the product of two random variables and asymptotically, the first of the random variables is a standard normal random  variable (by Equation~\ref{wn}), and the second is equal to the constant 1, 
the  Slutsky's theorem~\parencite[e.g.][p. 19]{s80} implies that asymptotically,
 $\tilde{\tilde{Z}}(\hat{\theta})$  follows the $\N(0,1)$ standard normal distribution.

\subsection*{An Alternative Variance Expression}
Note that by Equations~\ref{ssq} and for the expression of $c(\theta)$ provided in \ref{cexpr}, 
$\tau^2(\theta)$ can be expressed as
    \begin{eqnarray}
   \tau^2(\theta) &=& \frac{1}{n} \sum_i \tilde{w}^2_i(\theta) P_i(\theta)\{1- P_i(\theta)\} \nonumber\\ &=& \frac{1}{n} \sum_i \left\{w_i(\theta)-r_i(\theta)c(\theta)\right\}^2P_i(\theta)\{1- P_i(\theta)\} \nonumber\\
   &=& \frac{1}{n} \sum_i \left\{w_i(\theta)-\frac{r_i(\theta)}{I(\theta)}\sum_k P'_k(\theta) w_k(\theta) \right\}^2 P_i(\theta)\{1- P_i(\theta)\} \nonumber\\
   &=& \sigma^2(\theta) + \frac{1}{n} \frac{\left\{\sum_i P'_i(\theta) w_i(\theta)\right\}^2}{I^2(\theta)} \sum_i r^2_i(\theta)P_i(\theta)\{1- P_i(\theta)\} \nonumber \\ &-& \frac{2}{n}\frac{\sum_i P'_i(\theta) w_i(\theta)}{I(\theta)} \sum_i w_i(\theta) r_i(\theta) P_i(\theta)\{1- P_i(\theta)\} \cdot
 \label{tau2} 
  \end{eqnarray}
Using Equations~\ref{ri} and \ref{It}, the second term above is equal to 
\[ \frac{1}{n} \frac{\left\{\sum_i P'_i(\theta) w_i(\theta)\right\}^2}{I^2(\theta)} \sum_i \frac{ \{P'_i(\theta)\}^2}{P_i(\theta)\{1- P_i(\theta)\}} = \frac{1}{n} \frac{\left\{\sum_i P'_i(\theta) w_i(\theta)\right\}^2}{I(\theta)},  \] and the third term above is equal to 
\[ \frac{2}{n}\frac{\sum_i P'_i(\theta) w_i(\theta)}{I(\theta)} \sum_i w_i(\theta)  P'_i(\theta) = \frac{2}{n}\frac{\{\sum_i P'_i(\theta) w_i(\theta)\}^2}{I(\theta)} \]
Therefore, $\tau^2(\theta)$ can be expressed as 
\begin{eqnarray}
   \tau^2(\theta) 
   &=& \sigma^2(\theta) -  \frac{1}{nI(\theta)}\left \{\sum_i P'_i(\theta) w_i(\theta)\right\}^2 \cdot
   \label{tau3} 
    \end{eqnarray}
 \textcite{ljrv24} derived a  result like Equation~\ref{tau3} for the $l^*_z$ statistic.

  Because of Equation~\ref{thetaD}, for long tests,
  $\Var(\hat{\theta})$ is approximately equal to 
  $I^{-1}({\hat{\theta}})$. Therefore, an asymptotic approximation of
  $\tau^2(\hat{\theta})$ is obtained as
 \begin{eqnarray} \tau^2(\hat{\theta}) \approx
 \psi^2(\hat{\theta})  =  \sigma^2(\hat{\theta}) - \frac{1}{n}\widehat{\Var}(\hat{\theta})
 \left \{\sum_i P'_i(\hat{\theta})w_i(\hat{\theta})  \right \}^2,
 \label{psisq} 
 \end{eqnarray}
where $\widehat{\Var}(\hat{\theta})$ denotes an estimate of $\Var(\hat{\theta})$.
 Consequently, another expression of $\tilde{\tilde{Z}}(\hat{\theta})$ is given by
 \begin{eqnarray}
 \tilde{\tilde{Z}}(\hat{\theta})=\frac{W(\hat{\theta})}{\sqrt{n}\psi(\hat{\theta})} \cdot
 \label{zt2}
 \end{eqnarray}
 %

 \comment{
 Asymptotically, $\tilde{\tilde{Z}}(\hat{\theta})$ follows the standard normal distribution
 because
  \begin{itemize}
  \item
 \begin{eqnarray}
 \tilde{\tilde{Z}}(\hat{\theta})=\frac{W({\theta})}{\sqrt{n}\psi({\theta})}\times \frac{W(\hat{\theta})}{W({\theta})} \times
\frac {\sqrt{n}\psi({\theta})} {\sqrt{n}\psi(\hat{\theta})}
\label{z2}
 \end{eqnarray}
\item the first term in the right side of Equation~\ref{z2} follows the standard normal distribution asymptotically by the Lindeberg-Feller central limit theorem for independent random variables~\parencite[e.g.][p. 128]{r73}\footnote{because the terms of $W({\theta})$ are uniformly bounded by the regularity conditions and because $n\psi^2({\theta})\rightarrow \infty$, as in the proof of Lemma 1 of \textcite{s01}.}
\item  the second and third terms in the right side  converge to 1 because
$\hat{\theta}$ converges to ${\theta}$ and because of the result that $X_n \rightarrow c \Rightarrow g(X_n) \rightarrow g(c)$ for a sequence of random variables $X_n$ and any continuous function $g$~\parencite[e.g.][p. 124]{r73} 
 \item using Slutsky's theorem~\parencite[e.g.][p. 19]{s80},  because $\tilde{\tilde{Z}}(\hat{\theta})$ is the product of three random variables and asymptotically, the first of the random variables is a standard normal variable, and the latter two random variables  converge to the constant 1,
  $\tilde{\tilde{Z}}(\hat{\theta})$ follows the standard normal distribution. 
 \end{itemize}}
 \subsection*{The Ability Estimates for which the Asymptotic Normality of $\tilde{\tilde{Z}}(\hat{\theta})$  Holds}
It was mentioned earlier that the derivations in this paper underlie the assumptions that~(and hence the asymptotic normality of  $\tilde{\tilde{Z}}(\hat{\theta})$ holds when) $\hat{\theta}$ is the MLE or asymptotically equivalent to the MLE and satisfies the condition stated in 
Equation~\ref{thetaD}. The MLE of $\theta$ satisfies the condition stated in 
Equation~\ref{thetaD} for most common IRT models when it is reasonable to assume that the test is lengthened by adding strictly parallel forms~\parencite[e.g.][]{l83,w89,s15JEBS}.\footnote{\textcite{l83} stated that the latter condition is usually assumed to hold in IRT applications.}
So the asymptotic normality of  $\tilde{\tilde{Z}}(\hat{\theta})$ holds for the MLE.
A proof of the asymptotic equivalence of WLE and MLE under some regularity conditions\footnote{One condition is that the test is lengthened by adding strictly parallel forms.} is given in \textcite{w89}. 
The asymptotic equivalence of MLE and either of MAP or EAP, under some regularity conditions, can be inferred from the result of \textcite{cs93} that the asymptotic posterior distribution of ability is normal with mean equal to the MLE and a variance that converges to 0.\footnote{One of the regularity conditions of \textcite{cs93} is that the prior distribution is continuous and is positive at the true ability.} \textcite{cs93} noted that their regularity conditions hold for  all reasonable IRT models of well-designed tests.
Thus,  any asymptotic property of the MLE is shared by the WLE, MAP, and EAP and hence these estimates satisfy the condition stated in Equation~\ref{thetaD}. Appendix A includes the results of some simulations to demonstrate the asymptotic properties of MLE, WLE, MAP, and EAP. Consequently,
the asymptotic normality of  $\tilde{\tilde{Z}}(\hat{\theta})$ holds for WLE, MAP, and EAP as well.\footnote{Any discussion on
 the asymptotic  distribution of $\tilde{\tilde{Z}}(\hat{\theta})$ computed using the EAP or MAP makes sense when the latter two are treated as frequentist estimates (or as  functions of data from a frequentist viewpoint), just like  MAP was  treated as a frequentist estimate by \textcite{s01}.} 

 \subsection*{Relationship to the Results Provided in \textcite{s01}}
 Two questions, in light of the above derivations, are: ``What is the relationship between  $\tilde{\tilde{Z}}({\hat{\theta}})$ and  $\tilde{Z}({\hat{\theta}})$?'' and ``Does $\tilde{\tilde{Z}}({\hat{\theta}})$ have better or worse psychometric properties compared to  $\tilde{Z}({\hat{\theta}})$?'' 

 A comparison of  $\tilde{Z}({\hat{\theta}})$ and $\tilde{\tilde{Z}}({\hat{\theta}})$ provided in Equations~\ref{Wdis} and \ref{zt2},  respectively, indicates that their denominators are asymptotically equivalent by  Equation~\ref{psisq}; their
 numerators differ by the term $c({\hat{\theta}})r_0({\hat{\theta}})$, but, because $\frac{c({\hat{\theta}})r_0({\hat{\theta}})}{\sqrt{n}\tau(\hat{\theta})}\rightarrow 0$, as proved by \textcite[][p. 338]{s01}, the differences in their numerators divided by their common denominator converges to 0. Therefore, asymptotically,  $\tilde{Z}({\hat{\theta}})=\tilde{\tilde{Z}}({\hat{\theta}})$ and the derivations in this paper lead to the same standardized statistic as in \textcite{s01}. 
 \subsection*{Advantages of the Derivations of this Paper}
   Even though  $\tilde{Z}(\hat{\theta}) =\tilde{\tilde{Z}}(\hat{\theta})$ asymptotically,
   the derivations and expressions of this paper may offer some advantages over those of \textcite{s01}.
   
   First, unlike   the variance estimate $ \tau^2(\hat{\theta})$ suggested by  \textcite{s01}, the estimate $\psi^2(\hat{\theta})$ given in Equation~\ref{psisq} is approximately equal to  $\sigma^2(\hat{\theta})$
   minus a positive term for long tests, which  implies that the correct asymptotic
  variance estimate of  $\frac{1}{\sqrt{n}} W(\hat{\theta})$
 is smaller than $\sigma^2(\hat{\theta})$ that is used to compute statistics like $l_z$. 
  The magnitude of the reduction in variance is equal to  $\frac{1}{n}\widehat{\Var}(\hat{\theta})\left \{\sum_i P'_i(\hat{\theta})w_i(\hat{\theta})  \right \}^2$. Simulations by \textcite{s01} and others showed a reduction in variance of asymptotically correct standardized person-fit statistics for statistics like $l_z$. For example, \textcite{s01} showed using simulations that $ \tau^2(\hat{\theta})$   is always smaller than $ \sigma^2(\hat{\theta})$ for $l_z$. 
 Equation~\ref{psisq} provides a theoretical explanation of the result. \textcite{s01} also showed in his Table~1 that $ \tau^2(\hat{\theta})/\sigma^2(\hat{\theta})$ decreases as the underlying  $\theta$ increases from 0 to 2. An explanation of this result can be found from Equation~\ref{psisq}. As  $\theta$ increases from 0 to 2 (or, becomes more extreme), 
 \begin{itemize}
    \item $\widehat{\Var}(\hat{\theta})$ increases~\parencite[e.g.][]{kn93};
      \item  $P'_i(\hat{\theta})$, which is positive,\footnote{Because $P_i(\hat{\theta})$ is an increasing function of $\hat{\theta}$.}  does not change much;
    \item one is dealing with more able examinees for whom $w_i(\hat{\theta})$ tends to be positive for more items\footnote{because $P_i(\hat{\theta})$ tends to be larger than $(1-P_i(\hat{\theta}))$ for more items. For example, if the true difficulty parameters  are equi-spaced values between -2 and 2, the true slopes are all 1 and true guessing parameters are all 0 for a 50-item test, then $P_i(\hat{\theta})$ is larger than $(1-P_i(\hat{\theta}))$ for 25 items when $\hat{\theta}=0$, but for 49 items when $\hat{\theta}=2$.};
    \item $\sum_i P'_i(\hat{\theta})w_i(\hat{\theta})$ and its square become larger because of bullet points 2 and 3;
    \item  $ \frac{\psi^2(\hat{\theta})}{\sigma^2(\hat{\theta})}= \frac{\tau^2(\hat{\theta})}{\sigma^2(\hat{\theta})}$ decreases because of bullet points 1  and 4.
 \end{itemize}
 
  Second, the derivations of this paper promise to make the definition and/or description of the asymptotically correct standardized statistic considerably simpler, given that one does not have to define $c(\theta)$, $r_i(\theta)$, or $r_0(\theta)$ as in the description of the statistic in \textcite{s01}. 
  Table~\ref{diff} compares the definitions of the standardized statistic of \textcite{s01} and that suggested in this paper---the table makes it clear that the latter definition and/or description is  simpler.
  In addition, one does not need to bother about the regularity conditions associated with $r_i(\theta)$'s in using these statistics.\footnote{Two assumptions of Theorem 1 of Snijders, 2001 involve some conditions about the $r_i(\theta)$'s.} Also, Equation~\ref{T22} provides a clearer explanation of why the specific $\tilde{w}_i(\theta)$ 
   defined in \textcite{s01} works.\footnote{This is a somewhat important issue, mainly because \textcite{s01} did not derive the form of $\tilde{w}_i(\theta)$---he instead showed that setting  $\tilde{w}_i(\theta)=w_i(\theta)   -  c(\theta) r_i(\theta)$ leads to an asymptotically correct standardization.}
\begin{table}
\caption{The Difference in Definitions in Snijders (2001) and in this Paper.}
\begin{center}
\begin{tabular}{cc}
\hline  Snijders (2001) & This Paper \\ \hline
   The standardized statistic is given by & The standardized statistic is given by \\ 
$\frac{W(\hat{\theta})+c(\hat{\theta})r_0(\hat{\theta})}{ \sqrt{n}\tau(\hat{\theta})}$, & $\frac{W(\hat{\theta})}{\sqrt{n}\psi(\hat{\theta})}$,\\
  where $W(\theta) = \sum_{i=1}^{n} \left [ X_i - P_i(\theta)  \right ] w_i(\theta)$, & where $W(\theta) = \sum_{i=1}^{n} \left [ X_i - P_i(\theta)  \right ] w_i(\theta)$,\\
  $ \tau^2(\theta) = \frac{1}{n} \sum_i [\tilde{w}_i(\theta)]^2 P_i(\theta)[1- P_i(\theta)]$, & $ \psi^2(\theta) = \sigma^2(\theta) - \frac{1}{n}\Var(\hat{\theta})
 \left \{\sum_i P'_i(\theta)w_i(\theta)  \right \}^2 $,\\ 
  $\tilde{w}_i(\theta)= w_i(\theta) - c(\theta)r_i(\theta)$, & $ \sigma^2(\theta) = \frac{1}{n} \sum_i w^2_i(\theta) P_i(\theta)[1- P_i(\theta)]$, \\ 
 $r_i(\theta)=P'_i(\theta)/[P_i(\theta)(1-P_i(\theta))]$,  & $P'_i(\theta)$ = The first derivative of $P_i(\theta)$,\\ 
 $r_0(\hat{\theta})+\sum_{i=1}^{n} \left [ X_i - P_i(\hat{\theta})  \right ] r_i(\hat{\theta})$ = 0,& $\hat{\theta}$ is MLE, MAP, WLE, or EAP,\\
  $c(\theta) = \sum_i P'_i(\theta) w_i(\theta)/
 \sum_i P'_i(\theta) r_i(\theta)$, & and $\Var(\hat{\theta})$ = Variance of $\hat{\theta}$.\\
 $\hat{\theta}$ is MLE, MAP, or WLE,\\
  $ r_0(\hat{\theta}) =
 \left\{ \begin{array}{l} 0  \mbox{ if } \hat{\theta}=\mbox{MLE},
   \\ \frac{d
     \log f(\hat{\theta})}{d\hat{\theta}} \mbox{ if }
   \hat{\theta}=\mbox{MAP},
  \\ 
\frac{J(\hat{\theta})}{2I(\hat{\theta})} \mbox{ if }
 \hat{\theta}=\mbox{WLE},
      \end{array}  
\right .$ &\\
 $f(\theta)$ is the prior distribution on $\theta$, &\\
 $I(\theta)$ is the Fisher information on $\theta$, &\\
 $J(\theta)=\sum_i  \frac{P'_{i}(\theta) P''_{i}(\theta)}{P_{i}(\theta)(1-P_{i}(\theta))}$,&\\
 and $P''_{i}(\theta)$  is the second derivative of $P_{i}({\theta})$.  &\\
\hline
\end{tabular}
\label{diff}
\begin{center}
Note: Because \textcite{s01} only mentioned  MLE, MAP, and WLE, only those estimates are listed in the left column.
\end{center}
\end{center}
\end{table}
  
  Third,  no requirements involving $r_i(\theta)$ or $r_0(\theta)$ to compute $\tilde{\tilde{Z}}(\hat{\theta})$ make it  easier to apply the standardization suggested in this paper with ability estimates other than those considered in this paper, primarily because of an abundance of asymptotic results on the distribution of ability estimates~\parencite[e.g.,][]{cs93,l83,m14,m16,s15JEBS,w89}. 
  For example, it took a somewhat long proof~\parencite[by][]{s16BJ} to show that the standardization  of \textcite{s01} can be applied with the EAP, while it was straightforward~\parencite[because of asymptotic properties of EAP established by][]{cs93} to prove that the standardization derived in this paper can be applied with the EAP. 
  
  Finally, the derivations of this paper make it clear that the asymptotic mean of $W(\hat{\theta})$ is zero. There is some confusion about this mean in the paper of \textcite{s01}. While his Equation 21 implies that the asymptotic mean of $W(\hat{\theta})$ is zero (the result is proved in his Corollary 1), his Equation~12 specifies the mean to be $-c(\hat{\theta})r_0(\hat{\theta})$. Unfortunately, the latter  mean has been adopted in textbooks~\parencite[e.g.][p. 213]{d22}, software packages~\parencite[e.g.][]{tmn16}, and journal articles~\parencite[e.g.][]{vm99}.
  While the incorrect inclusion of  
  $-c(\hat{\theta})r_0(\hat{\theta})$ in the mean will not matter if the MLE is used (because $r_0(\hat{\theta})=0$ for MLEs, which makes the mean equal to 0 even with the incorrect formula) and is not likely to lead to much of a difference for long tests for other ability estimators (because $-c(\hat{\theta})r_0(\hat{\theta})/{\sqrt{n}\tau(\hat{\theta})}$ is approximately 0 for long tests), it may matter for short tests, as is demonstrated using simulations below.
 \bas{1}
\begin{table}[h!tb]
\caption{Type I Error Rates of the Person-fit Statistics.}
\begin{center}
\begin{tabular}{cccccccccc}
\hline
 Statistic \&  & \multicolumn{3}{c}{Test Length=12}&\multicolumn{3}{c}{Test Length=36}&\multicolumn{3}{c}{Test Length=72}\\ \cline{2-10}
 $\theta$-estimate& $\alpha$=.01 & $\alpha$=.02& $\alpha$=.05 & $\alpha$=.01 & $\alpha$=.02& $\alpha$=.05 & $\alpha$=.01 & $\alpha$=.02& $\alpha$=.05    \\\hline
    $l^*_z$\&MLE &  .017 & .029 & .058 & .015 & .025 & .052 & .013 & .023 & .050\\
   $l^*_z$\&WLE &  .016 & .027 & .055 & .014 & .024 & .051 & .013 & .022 & .050\\
   $l^{**}_z$\&WLE & .012 & .021 &.042 &.012 & .020 & .044 & .011 & .020 & .045 \\
$l^{**}_z$\&EAP & .012 & .020 & .041 &.011 & .020 & .044 & .011 & .020 & .044 \\   
\hline
\end{tabular}
\label{T1}
\end{center}
\end{table}
\bas{2}
  Table~\ref{T1} shows the Type~I error rates at 1\%, 2\%, and 5\% levels for some simulated data for
  \begin{itemize}
 \item $l^*_z$ computed using the R package PerFit~\parencite{tmn16} using MLE and WLE 
 \item $l^{**}_z$, which is computed using an R program written by the author to compute $\tilde{\tilde{Z}}(\hat{\theta})$, computed with WLE and EAP and using $w_i(\theta)=\log\frac{P_i(\theta)}{1-P_i(\theta)}$. 
  \end{itemize}
  While $l^*_z$ with MLE does not involve any mean correction,\footnote{Which is why $l^{*}_z$ with MLE is identical to $l^{**}_z$ with MLE.}  $l^*_z$ with WLE is computed (in R package PerFit) under the assumption that the  asymptotic mean of $W(\hat{\theta})$ is $-c(\hat{\theta})r_0(\hat{\theta})$, and  $l^{**}_z$ with WLE and EAP are computed under the assumption that the  asymptotic mean of $W(\hat{\theta})$ is 0. The Type~I error rates were computed
  from  data sets simulated from the 3PL model. Three test lengths---12, 36, and 72---were considered. The true ability was assumed to follow the standard normal distribution.
  For each test length, the Type~I error rate at each level was computed from 1,000 data sets (with 1,000 examinees each) that involved no person misfit. 
  The table shows that the Type~I error rates of $l^*_z$ are often inflated, both for the MLE and WLE---especially for shorter tests and smaller significance levels---while those of $l^{**}_z$ are close to and often just below the nominal level. 
  Given the popularity of person-fit statistics in detecting test
  fraud~\parencite[e.g.][]{ws18}, the recommendation by several experts~\parencite[e.g.][]{wce15} of conservative hypothesis testing in test-fraud investigations, 
  and the severe consequences of a false alarm in a test-fraud investigation~\parencite[e.g.][]{w16}, many practitioners would prefer $l^{**}_z$ with WLE or EAP over $l^*_z$ with either ability estimate, especially for tests that are not too long. Thus, the assumption that the approximate asymptotic mean of $W(\hat{\theta})$ is $-c(\hat{\theta})r_0(\hat{\theta})$ may matter in practice. 

  One limitation of $\tilde{\tilde{Z}}(\hat{\theta})$ is that it is unclear if its
  asymptotic normality  holds when the statistic is computed using robust ability estimates~\parencite[e.g.][]{mb82,sy11}, mainly because the variance of a robust estimate does not seem to be asymptotically equal to the inverse of the Fisher information.
  However, Appendix~B includes a proof that a statistic that looks slightly more complicated than  $\tilde{\tilde{Z}}(\hat{\theta})$ has an asymptotic standard 
  normal distribution when it is computed using a robust ability estimate.

 %
 \section*{Conclusions}
\label{conc}
 This paper provides an alternative derivation
 of the asymptotic standardization that \textcite{s01} suggested for a broad class of person-fit statistics. 
 While the derivations do not yield a new statistic, they offer a more direct route to the same asymptotic result and may therefore be easier to understand than the derivations of \textcite{s01}. In addition, the approach led to a simpler description of the asymptotic standardization, as demonstrated in Table~\ref{diff}, and to a theoretical explanation of simulation results reported by researchers like \textcite{s01}.
 Another potentially useful contribution of this paper is the demonstration that the asymptotic mean correction of $-c(\hat{\theta})r_0(\hat{\theta})$
 mentioned in \textcite{s01} and adopted in several publications is in fact unnecessary, and the correction may lead to inflated Type~I error rates.

 The class of person-fit statistics that this paper focused on is appropriate
 when the interest lies in testing against an unspecified general alternative
 \parencite[e.g.][p. 175]{gd07} and may not be the most
 appropriate for a particular problem. 
 If the anticipated model violation
 is more specific, a statistic such as the
 modified signed likelihood ratio statistic \parencite{sj19} may be more powerful. 
 In addition,  person-fit statistics should be used in
conjunction with other sources of information---such as seating charts, video surveillance,
or follow-up interviews---rather than as stand-alone indicators of test fraud~\parencite[e.g.][]{bn80,hhb87,mees08,tm14}.

 Several limitations of this paper suggest directions for future research.
 First,  this paper focused only on the standardized statistics of the type suggested in \textcite{s01}, whereas other person-fit statistics with known null distributions are available~\parencite{k95,s16JEBS}.  
 Second, as in \textcite{s01}, the derivations of this paper rely on several assumptions. Further work could investigate if and when those assumptions do not hold and the consequences on the results. Third, it may be worthwhile to examine if the derivations hold for ability estimates not considered in this paper. Finally, various versions of Snijders's approach  have been suggested for polytomous items \parencite{s16aa}, unfolding models \parencite{t16}, multidimensional IRT models \parencite{amt16,hlc21}, and the Rasch testlet model \parencite{ljrv24}---future research could focus on applying the types of derivations of this paper to simplify those versions.
\raggedright
\printbibliography
\newpage
\section*{Appendix A: Some Simulation Results Demonstrating the Asymptotic Properties of Four Ability Estimates}
\setlength{\topmargin}{0in} \pagenumbering{arabic} \setcounter{page}{1}
\setcounter{figure}{0} 
\renewcommand\thefigure{A\arabic{figure}}
\renewcommand{\theequation}{A\arabic{equation}}
\setcounter{equation}{0} 
\quad \quad 
A simulation study was performed to illustrate the consistency and asymptotic normality
of four ability estimates---MLE, WLE, MAP, and EAP. The similations involve five examinees with
true abilities ($\theta$) of  -2,  -1, 0, 1, and 2 and
four test lengths, with 12, 60, 240, and 1,200 dichotomous items. The two-parameter logistic (2PL) model was used in the simulations.
For the 12-item test:
\begin{itemize}
\item The true slope parameters of all items were generated from a
log-normal distribution with mean 
and standard deviation (SD) of the logarithm of the variable equal to 0 and 0.2, respectively.
\item The true difficulty  parameters for the 2PL model were generated from a standard normal distribution.
\end{itemize}

\quad \quad
 The item parameters of the longer tests were  replications (5, 20, or 100 times) of the item parameters of the 12-item test.  This strategy was used to  ensure the regularity condition that the test is lengthened by adding strictly parallel forms. 
 The true item parameters were generated only once and then used in all replications and all examinees for computing the ability estimates.
 For each examinee, 100,000 sets/replications of item scores
were generated for each test using the true ability and the true item parameters. 
 For each examinee and each replication, the MLE, WLE, MAP, and EAP were computed using the R package {\it PP}~\parencite{rs26}. The MLEs were forced to have a maximum absolute value of 4 (thus, e.g., the MLE for an examinee with $X_i=0$ for all $i$ was set equal to -4). The WLE, MAP, and EAP were always finite. The MAP and EAP were computed using a standard normal prior distribution.

 \begin{figure}[h!tb]
\centerline{\epsfig{figure=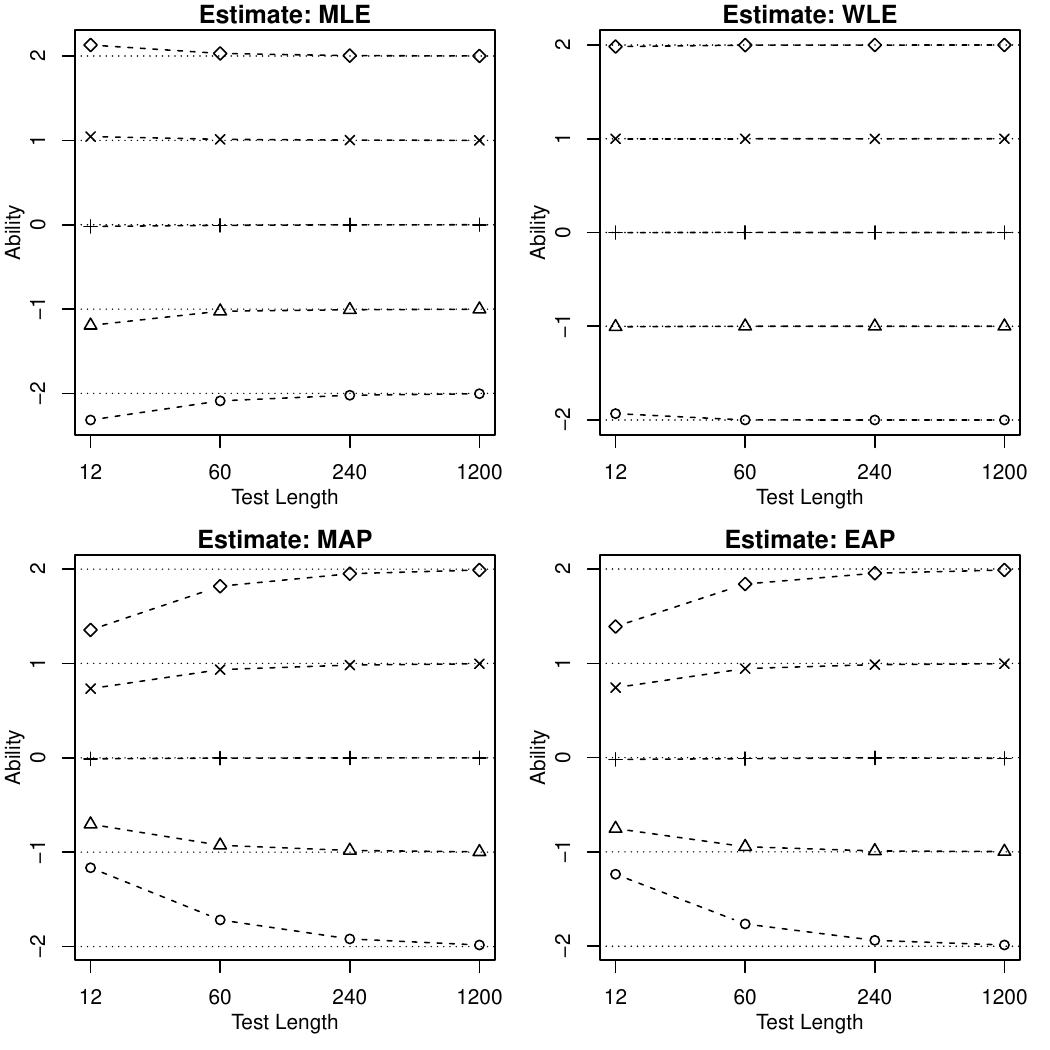,width=6.5in,height=6.5in}}
 \caption{Average of the ability estimates for increasing test lengths for five examinees.}
  \label{consis} 
\end{figure}
\quad \quad Each panel of Figure~\ref{consis} shows, for one ability estimate, the average estimate~(average computed from 100,000 replications) for the four test lengths for the five examinees. The four averages for an examinee, plotted using the same symbol, were connected using a dashed line. The five symbols---diamond, cross, plus, triangle, and circle---correspond to the examinees with true abilities of 2, 1, 0, -1, and -2, respectively.
Five horizontal dotted lines in each panel indicate the true examinee abilities. 
The figure shows that the averages are often visibly different from the corresponding true values for shorter test lengths, but converge to the true values as test length increases. Convergence occurs much faster for MLE and WLE compared to the MAP and EAP, especially for true abilities different from 0. An interesting finding is that for EAP and MAP, even the average estimates for the examinees with true abilities -2 or 2 are essentially equal to the corresponding true values for longer tests even though the mean of the prior distribution, 0, is far from the true values. Overall, Figure~\ref{consis} indicates that all the four ability estimates are asymptotically equivalent, although the speed of convergence varies over the estimates.

\comment{
\begin{figure}[h!tb]
\centerline{\epsfig{figure=wle.pdf,width=6.5in,height=6.5in}}
 \caption{Distribution of the WLE for increasing test lengths for five examinees.}
  \label{dWLE} 
\end{figure}
\quad \quad 
Each row of Figure~\ref{dWLE} shows the kernel-smoothed distributions (smoothed
using the function ‘‘density’’ in the R software package) of the WLEs for one 
examinee for each of the five tests. The distribution plotted in each panel is based on 10,000 WLEs computed for an examinee for the corresponding test. In any panel, a vertical dashed line shows the true 
ability of the examinee, and the title shows the test length and true examinee ability.
A figure for the MLEs looks very similar to  Figure~\ref{dWLE}. 
Figure~\ref{dMAP} shows same plots for the MAP estimates for the five examinees. A figure for the EAPs looks very similar to Figure~\ref{dMAP}.
\begin{figure}[h!tb]
\centerline{\epsfig{figure=map.pdf,width=6.5in,height=6.5in}}
 \caption{Distribution of the MAP estimate for increasing test lengths for five examinees.}
  \label{dMAP} 
\end{figure}

\quad \quad
Figures~\ref{dWLE} and \ref{dMAP}  show that even though the distribution of either ability estimate appears far from normal for shorter tests, the distribution looks much closer to a normal distribution as test length increases. In addition, as test length increases, the variance or width of the distribution becomes negligible and the center of the distribution becomes closer to the true value of the examinee ability.
These figures~(and similar figures for MLE and EAP) indicate that all of these estimates converge to a normal distribution with mean equal to the true value and variance that converges to zero.} 

\begin{figure}[h!tb]
\centerline{\epsfig{figure=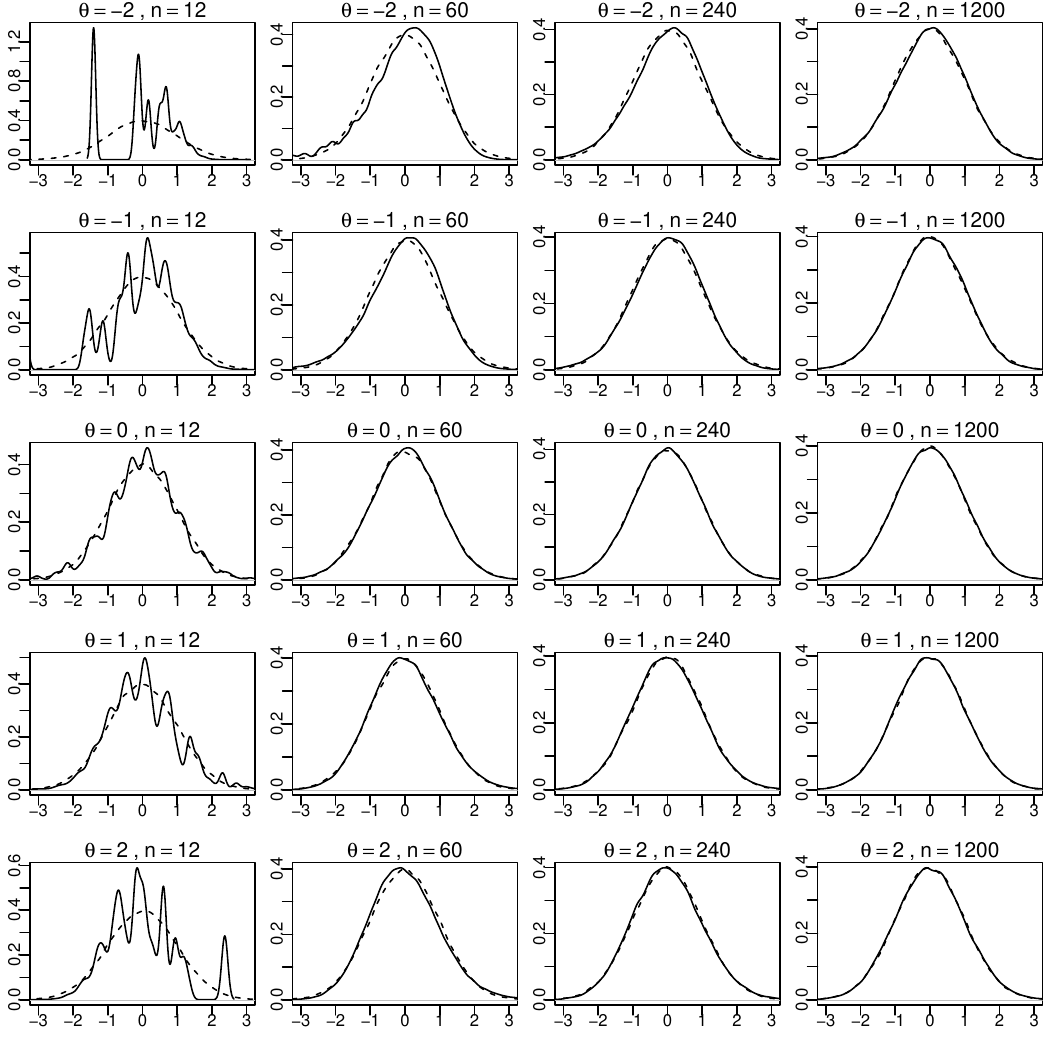,width=6.5in,height=6.5in}}
 \caption{Distribution of  $\sqrt{n\bar{I}_0(\theta)}$(WLE-$\theta$)  for increasing test lengths for five examinees.}
  \label{dWLE} 
\end{figure}
\quad \quad Each row of Figure~\ref{dWLE} shows the kernel-smoothed distributions (smoothed using the function ‘‘density’’ in the R software package), using solid lines, of 
$\sqrt{n\bar{I}_0(\theta)}$(WLE-$\theta) $, where $\theta$ is the true ability,
for one examinee for the four test lengths. Because the longer tests were constructed by replicating the original 12 items~(that constituted the starting point in the simulation), for any examinee, $\bar{I}_0(\theta)=I(\theta)/n$ was computed using Equation~\ref{It} and the item parameters of those 12 items and the examinee's true ability. 
The distribution plotted in each panel is based on 100,000 WLEs computed for an examinee for the corresponding test length. In each panel, a dashed line shows the density of the standard normal distribution and the title shows the true examinee ability and test length~($n$).
A figure for the MLEs looks very similar to  Figure~\ref{dWLE} and is not shown here. 
Figure~\ref{dMAP} shows similar plots for the MAP estimates for the five examinees. A figure for the EAPs looks very similar to Figure~\ref{dMAP}  and is not shown here.
\begin{figure}[h!tb]
\centerline{\epsfig{figure=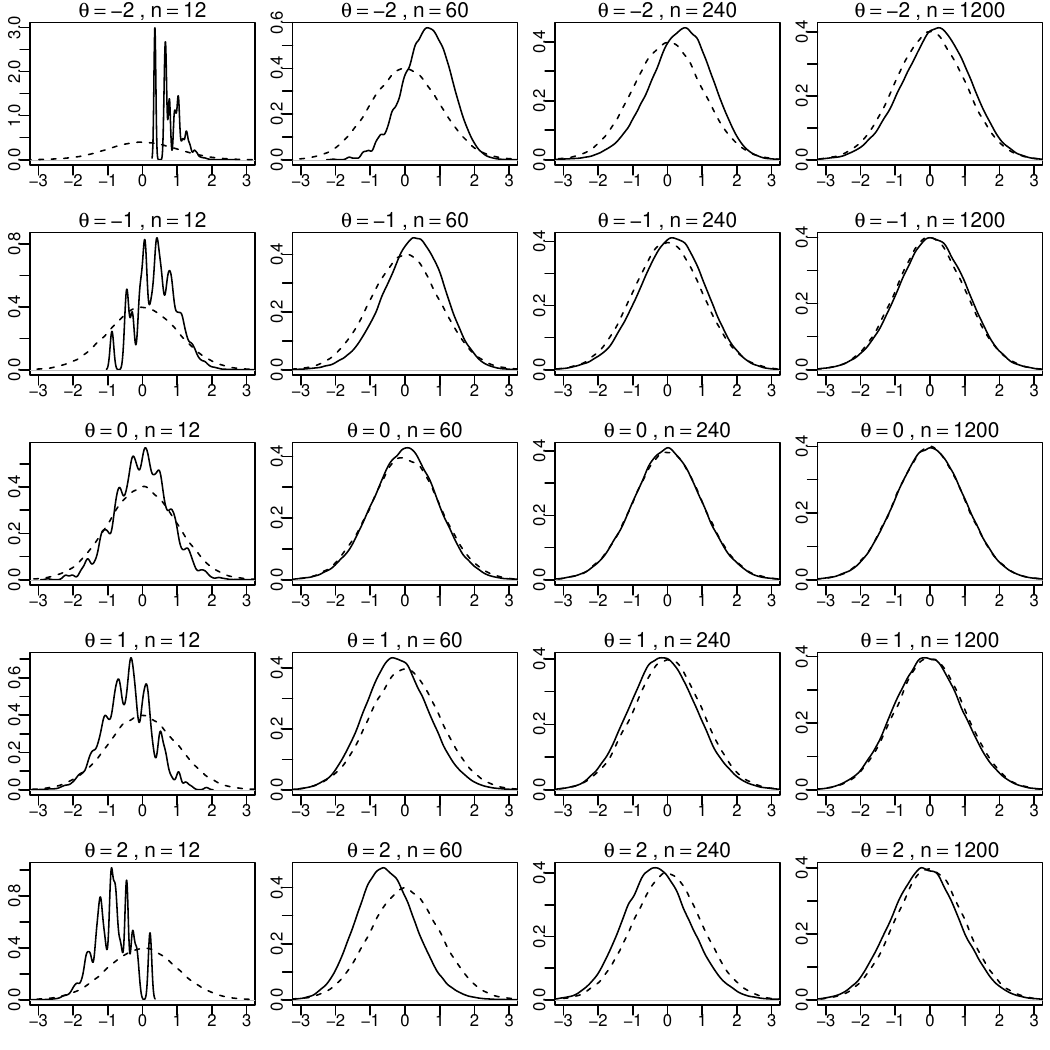,width=6.5in,height=6.5in}}
 \caption{Distribution of $\sqrt{n\bar{I}_0(\theta)}$(MAP-$\theta) $ 
 for increasing test lengths for five examinees.}
  \label{dMAP} 
\end{figure}

\quad \quad
Figures~\ref{dWLE} and \ref{dMAP}  show that even though the distribution for either ability estimate appears far from the standard normal distribution and the mean is far from 0 for shorter tests~(more so for the MAP), the distribution looks much closer to a  standard normal distribution. These figures~(and similar figures for MLE and EAP) indicate that 
all of these estimates satisfy the condition provided in Equation~\ref{thetaD}.\footnote{These figures show that $\sqrt{n\bar{I}_0({\theta})}(\hat{\theta}-\theta) \indist \N(0,1)$, or, that $\sqrt{n}(\hat{\theta}-\theta) \indist \N(0,\bar{I}^{-1}_0({\theta}))$ for longer tests, where $\hat{\theta}$ is the MLE, WLE, MAP, or EAP for each of the five values of $\theta$ considered here.}

\quad \quad
Figures~\ref{consis} to \ref{dMAP} together seem to indicate that the two conditions required from the ability estimates for the derivations of this paper to hold are satisfied for MLE, WLE, MAP, and EAP.

\newpage
\section*{Appendix B: Asymptotically Correct Standardization for Robust Estimates}
\setlength{\topmargin}{0in} \pagenumbering{arabic} \setcounter{page}{1}
\setcounter{figure}{0} 
\renewcommand\thefigure{B\arabic{figure}}
\renewcommand{\theequation}{B\arabic{equation}}
\setcounter{equation}{0} 
\quad 
 Robust estimates including the biweight~\parencite{mb82} and Huber estimate~\parencite{sy11} are computed by solving for $\theta$ the equation 
 \begin{eqnarray}
 \sum_i [ X_i - P_i(\theta) ] 
 \frac{\zeta_i(\theta)P'_i(\theta)}{P_i(\theta)\{1-P_i(\theta)\}} 
\label{robeqn}
 \end{eqnarray}
  ~\parencite[e.g.][]{m16}. The function $\zeta_i(\theta)$ is defined as 
  $\zeta_i(\theta)=\eta(a_i(\theta-b_i))$, where
  \begin{eqnarray*}
 \eta(t)=
 \left\{ \begin{array}{l} (1-t^2/16)^2 \mbox{ if } |t|\le 4,\\ 0 \mbox{ otherwise,}
      \end{array}
\right .
\end{eqnarray*}
for the biweight estimate~\parencite{mb82} and as 
\begin{eqnarray*}
 \eta(t)=
 \left\{ \begin{array}{l} 1 \mbox{ if } |t|\le 1,\\ 1/|t| \mbox{ otherwise,}
      \end{array}
\right .
\end{eqnarray*}
for the Huber estimate~\parencite{sy11}.
 These estimates, denoted as $\hat{\theta}$, can be approximated as 
 \begin{eqnarray}
  \hat{\theta} \approx \theta + \frac{\sum_i \{\alpha_i(\theta)X_i+ \beta_i(\theta)\}} {\sum_i\zeta_i(\theta)I_i(\theta)}, 
 \label{thapp}
 \end{eqnarray}
 where \[ \alpha_i(\theta)=\frac{\zeta_i(\theta)P'_i(\theta)}{{P_i(\theta)\{1-P_i(\theta) \}}}, \beta_i(\theta)=\frac{\zeta_i(\theta)P_i(\theta)}{1-P_i(\theta)},
 I_i(\theta)=\frac{(P'_i(\theta))^2}{P_i(\theta)\{1-P_i(\theta)\}}, \] because of results provided in Equation 42 and Table~1 of \textcite{m16}.
 Consequently, 
 \begin{eqnarray}
 \Cov(X_i,\hat{\theta})\approx \frac{\alpha_i(\theta)}{\sum_i\zeta_i(\theta)I_i(\theta)}\Var(X_i) =  \frac{\zeta_i(\theta)P'_i(\theta)}{\sum_i\zeta_i(\theta)I_i(\theta)} \cdot
 \end{eqnarray}
 Some further derivations lead to the result that
 \begin{eqnarray}
 \tilde{\psi}^2(\theta)  \approx    \sigma^2(\theta) + \frac{1}{n}\Var(\hat{\theta})
 \left \{\sum_i P'_i(\theta)w_i(\theta)  \right \}^2 -\frac{2}{n}\frac{\sum_i P'_i(\theta)w_i(\theta) \sum_i P'_i(\theta)w_i(\theta) \zeta_i(\theta) }{\sum_i\zeta_i(\theta)I_i(\theta)} \cdot
 \label{TA2} 
 \end{eqnarray}
 Then, it can be proved, in a manner similar to that above, that the asymptotic null distribution of
 \begin{eqnarray}
 \tilde{\tilde{\tilde{Z}}}(\hat{\theta})=\frac{W(\hat{\theta})}{\sqrt{n}\tilde{\psi}(\hat{\theta})}
 \end{eqnarray}
is standard normal.

\quad\quad
 Equation~5 of \textcite{s01} and Equation~\ref{robeqn} above imply that for the robust estimates, 
 \begin{eqnarray}
 r_i(\theta)=\frac{\zeta_i(\theta)P'_i(\theta)}{P_i(\theta)\{1-P_i(\theta)\}},
 \label{rirob}
 \end{eqnarray}
a result that was proved by \textcite{s16BJ}. 
 Using the expression of  $r_i(\theta)$ in Equation~\ref{rirob},  the $c(\theta)$ given in Equation~\ref{cn} and in the derivations of \textcite{s01} is given by  
  \begin{eqnarray}
 c(\theta)=\frac{\sum_i P'_i(\theta) w_i(\theta)}
 {\sum_i P'_i(\theta)r_i(\theta)}=  \frac{\sum_i P'_i(\theta) w_i(\theta)}
 {\sum_i \zeta_i(\theta)\{P'_i(\theta)\}^2 / [ P_i(\theta)\{1- P_i(\theta)\}]} = \frac{\sum_i P'_i(\theta) w_i(\theta)}{\sum_i \zeta_i(\theta)I_i(\theta)} \cdot
 \label{capp}
  \end{eqnarray}
 Then the variance provided by \textcite{s01} and given in Equation~\ref{tau} can be expressed as
  \begin{eqnarray}
   \tau^2(\theta) &=& \frac{1}{n} \sum_i \left\{w_i(\theta)-r_i(\theta)c(\theta)\right\}^2P_i(\theta)\{1- P_i(\theta)\}  \nonumber\\
   &=& \sigma^2(\theta) + c^2(\theta)\frac{1}{n} \sum_i r^2_i(\theta)P_i(\theta)\{1- P_i(\theta)\} -  \frac{2c(\theta)}{n} \sum_i r_i(\theta) w_i(\theta) P_i(\theta)\{1- P_i(\theta)\} \nonumber\\
   &=& \sigma^2(\theta) + c^2(\theta)\frac{1}{n} \sum_i \frac{\zeta^2_i(\theta)\{P'_i(\theta)\}^2}{P_i(\theta)\{1- P_i(\theta)\}} -  \frac{2c(\theta)}{n} \sum_i w_i(\theta) \zeta_i(\theta)  P'_i(\theta) \nonumber \\
   &=& \sigma^2(\theta) + c^2(\theta)\frac{1}{n} \sum_i \zeta^2_i(\theta)I_i(\theta) -  \frac{2c(\theta)}{n} \sum_i w_i(\theta) \zeta_i(\theta)  P'_i(\theta)
   \cdot
 \label{tau2r} 
  \end{eqnarray}
  Equations~\ref{capp} and \ref{tau2r}  imply that
  \begin{eqnarray}
   \tau^2(\theta) = \sigma^2(\theta) &+& \frac{ \left \{\sum_i P'_i(\theta)w_i(\theta)  \right \}^2 \sum_i \zeta^2_i(\theta)I_i(\theta)}
   {n \{\sum_i \zeta_i(\theta)I_i(\theta) \}^2} \nonumber\\
&-&\frac{2}{n}\frac{\sum_i P'_i(\theta)w_i(\theta) \sum_i P'_i(\theta) w_i(\theta) \zeta_i(\theta) }{\sum_i\zeta_i(\theta)I_i(\theta)} \cdot
   \label{tau3r} 
  \end{eqnarray}
 Because the variance of the robust estimate is given by 
  \[ \Var(\hat{\theta}) = \frac{\sum_i \zeta^2_i(\theta)I_i(\theta)}{\{\sum_i \zeta_i(\theta)I_i(\theta) \}^2}\]
   ~\parencite[e.g.][Equation 21]{m16}, the variance given by Equation~\ref{tau3r} is equal to that given by Equation~\ref{TA2}. Thus,
   $\tilde{\tilde{\tilde{Z}}}(\hat{\theta})=\tilde{Z}(\hat{\theta})$  for robust ability estimates.
\end{document}